\documentclass[pra,twocolumn,superscriptaddress]{revtex4-2}

\usepackage{graphicx,subfig}
\usepackage{dcolumn}
\usepackage{bm,bbm}
\usepackage{amssymb,amsmath,amsthm}
\usepackage{color}
\usepackage{soul}
\usepackage[bookmarks=false]{hyperref}
\hypersetup{colorlinks=true,citecolor=blue,linkcolor=red,%
	urlcolor=blue,pdfstartview=FitH,bookmarksopen=true}
\makeatother
\usepackage{csquotes}
\MakeOuterQuote{"}

\newcommand{\Rmnum}[1]{\expandafter\@slowromancap\romannumeral #1@}

\newcommand{\kt}[1]{\vert #1 \rangle}
\newcommand{\ba}[1]{\langle #1 \vert}

\newcommand{\AffiBAQIS}{\affiliation{Beijing Academy of Quantum Information Sciences, Beijing 100193, China}}
\newcommand{\AffiBUPT}{\affiliation{School of Science and State Key Laboratory of Information Photonics and Optical Communications, Beijing University of Posts and Telecommunications, Beijing 100876, China}}

\begin{document}
	
	\title{Controlled entanglement source for quantum cryptography}
	
	\author{Qiang Zeng}	\email{zengqiang@baqis.ac.cn} \AffiBAQIS
	\author{Haoyang Wang} \AffiBAQIS \AffiBUPT
	\author{Huihong Yuan} \AffiBAQIS
	\author{Yuanbin Fan} \AffiBAQIS
	\author{Lai Zhou} \AffiBAQIS
	\author{Yuanfei Gao} \AffiBAQIS
	\author{Haiqiang Ma} \AffiBUPT
	\author{Zhiliang Yuan}	\email{yuanzl@baqis.ac.cn}	\AffiBAQIS
	
	\date{\today}
	
	\begin{abstract}
	Quantum entanglement has become an essential resource in quantum information processing.
	Existing works employ entangled quantum states to perform various tasks, while little attention is paid to the control of the resource.
	In this work, we propose a simple protocol to upgrade an entanglement source with access control through phase randomization at the optical pump.
	The enhanced source can effectively control all users in utilizing the entanglement resource to implement quantum cryptography.
	In addition, we show this control can act as a practical countermeasure against memory attack on device-independent quantum key distribution at a negligible cost.
	To demonstrate the feasibility of our protocol, we implement an experimental setup using just off-the-shelf components and characterize its performance accordingly.
	\end{abstract}
	
	\maketitle
		
\section{Introduction}\label{Sec:Intro}
	Modern society operates upon reliable and secure communication.
	Current cryptosystems defend from eavesdropping and attacks based on computation complexity, which is however vulnerable to advancing quantum technologies~\cite{xu2020Secure}.
	Quantum key distribution (QKD)~\cite{BB84,E91,lo2005Decoy,wang2005beating,MDI2012,TFQKD2018} enables distant users to establish secret keys assuming any computational power of attackers but rather based on the basic physical principles, thus guaranteeing information-theoretic security~\cite{rennerInformationtheoreticSecurityProof2005}.
	In particular, entanglement-based QKD (EB QKD) protocol~\cite{E91,BBM92} stands out compared to QKD using weak coherent pulses~\cite{fan-yuanRobustAdaptableQuantum2022,wangTwinfieldQuantumKey2022,zhouTwinfieldQuantumKey2023} for it requiring no intensity modulation~\cite{lo2005Decoy,wang2005beating} and is naturally applicable for multiusers~\cite{fitzkeScalableNetworkSimultaneous2022,kimQuantumCommunicationTimebin2022,neumannContinuousEntanglementDistribution2022,joshiTrustedNodeFree2020,wengerowskyEntanglementbasedWavelengthmultiplexedQuantum2018}.
	Further, the ultimate form of quantum secure communication~\cite{zhangDeviceindependentQuantumKey2022}, device-independent QKD protocol (DI QKD)~\cite{mayersQuantumCryptographyImperfect1998,barrettNoSignalingQuantum2005}, exploits the strong correlation existing in quantum entanglement that is beyond classical regime to eliminate fully the need of trust on the QKD users' apparatuses.
	The main insight behind this is that the security in DI QKD requires the exclusion of existence of the local hidden variables (LHV) model~\cite{brunner2014Bell} of the apparatuses, while such a model can be regarded as some preset conspiracies which cannot be distinguished in traditional QKD~\cite{acinBellTheoremSecure2006}.
	
	Indeed, a lower requirement on the level of trust comes with extra demanding requirements on the performance of the practical devices.
	To faithfully implement DI QKD, a high-quality entanglement resource is prerequisite to defend against collective attack~\cite{PhysRevLett.98.230501}. 
	In addition, high detection efficiency from the resource to both of the users is required to ensure an unbiased statistics to defend against detection efficiency attack~\cite{gisin1999LHV,hoNoisyPreprocessingFacilitates2020}.
	Moreover, the measurement device should be \emph{memoryless} to prevent from leaking information by tracing the history statistics of the reused devices, which is described as memory attack~\cite{barrettMemoryAttacksDeviceIndependent2013}.
	Fortunately, recent progresses show the above problems can be well tackled to meet the capability of state-of-the-art equipment.
	Remarkably, three practical DI QKD experiments were demonstrated in recent time~\cite{liuPhotonicDemonstrationDeviceIndependent2022,nadlingerExperimentalQuantumKey2022,zhangDeviceindependentQuantumKey2022}.
	
\begin{figure}[ht]
		\includegraphics[width=\columnwidth]{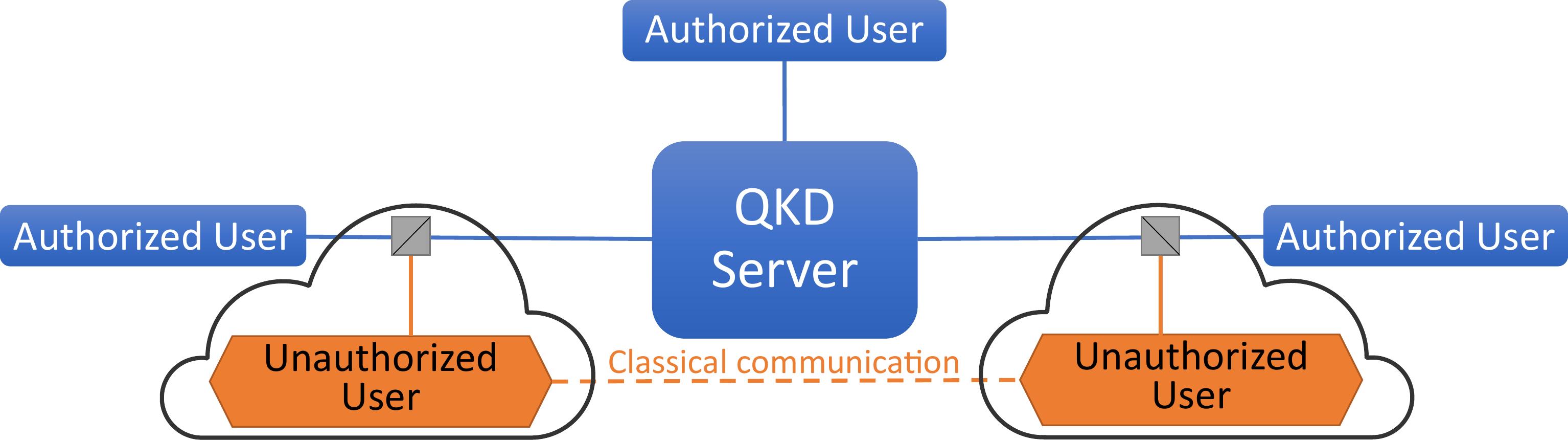}
		\caption{\label{F1}
			Schematic of entanglement-based QKD network with both authorized and unauthorized users.}
	\end{figure}

	Quantum entanglement is foreseeable to serve as a useful resource in future networks as a quantum utility just like electricity and gas in today's energy grids.
	For instance, the dealers distribute entangled states---usually carried with photon, as it is the best information carrier over long distance---to users to establish secret keys.
	However, the topic of resource control has rarely been discussed despite its apparent importance.
	As illustrated in Fig.~\ref{F1}, a conventional entanglement server constantly distributes maximally entangled states to the communication user pairs.
	However, the server lacks the ability to control whether the distributed photons are indeed consumed by authorized users.
	Unauthorized users can tap entangled photons off the channels for their own key generation without being noticed.
	
	Surprisingly, adding control to an entanglement source can also provide defense against memory attacks~\cite{barrettMemoryAttacksDeviceIndependent2013} for DI QKD, where the attackers can retrieve the secure keys generated in earlier sessions without being detected when the measurement devices are reused. 
	Existing countermeasures~\cite{barrettMemoryAttacksDeviceIndependent2013,curtyFoilingCovertChannels2019}, including destroying the used measurement devices, strictly isolating the entanglement generation devices or using additional alternate devices, essentially prevent the attackers' access to the (possibly) recorded information, but bring considerable complexity to the system, not to mention the prohibitive cost.
	
	Quantum mechanics provides an elegant solution to the control problems via quantum secret sharing (QSS)~\cite{zukowski1998Quest,PhysRevLett.83.648}.
	Its implementation, which originally required exotic multiphoton Greenberg-Horne-Zeilinger (GHZ) states~\cite{hillery1999Quantum,chen2005Experimental}, has been substantially simplified through the use of \emph{pseudo}-GHZ states comprising just photon pairs and thus enabled a QSS demonstration based on time-bin entanglement~\cite{tittel2001Experimentala}. 
	Recently, the two-photon QSS approach was successfully extended to polarization encoding~\cite{williams2019Quantum}.
	
	Inspired by the two-photon QSS protocol~\cite{tittel2001Experimentala}, we propose a simple but profoundly useful upgrade to a conventional entanglement source to have access control in which a binary pseudo randomness is introduced.
	By further introducing a mixed state, we show this enhanced entanglement source can effectively protect DI QKD from memory attacks at a negligible cost.
	We experimentally demonstrate our scheme in a time-bin encoding system.
	
\section{phase modulation to entanglement source}\label{Sec:mod1}
	We first briefly review the three-party QSS protocol.
	Assuming a triple-qubit GHZ state $\kt{\Psi}_{\text{GHZ}}=\frac{1}{\sqrt{2}}(\kt{000}+\kt{111})$ is shared by three parties (say, Alice, Bob, and Coy).
	Subjected to Alice performing measurement $\sigma_{\textsf{X}}$ or $\sigma_{\textsf{Y}}$ locally, where $\sigma_{\textsf{X}}$ and $\sigma_{\textsf{Y}}$ are the Pauli matrices, the state shared between the other two parties is projected onto one of the following four possible states, based on the outcome of Alice's local measurement:
	$$
	\begin{aligned}
		\textsf{X}^{\pm}_A&\rightarrowtail \kt{\phi^{\pm}}_{BC}=\frac{1}{\sqrt{2}}(\kt{00}\pm \kt{11}),\\
		\textsf{Y}^{\pm}_A&\rightarrowtail \kt{\psi^{\mp}}_{BC}=\frac{1}{\sqrt{2}}(\kt{00}\mp i \kt{11}),
	\end{aligned}
	$$
	where $\textsf{X}^{\pm}=\frac{1}{\sqrt{2}}(\kt{0}\pm\kt{1})$,
	$\textsf{Y}^{\pm}=\frac{1}{\sqrt{2}}(\kt{0}\pm i \kt{1})$ are the eigenstates of $\sigma_{\textsf{X}}$ and $\sigma_{\textsf{Y}}$, respectively.
	Alice's measurement outcome is thus required for Bob and Coy if they want to perform the conventional EB QKD protocol, and in this sense Alice can be regarded as the QKD controller, which is also known as third-man quantum cryptography~\cite{chen2005Experimental}.
	Instead of using a genuine triple-qubit GHZ state, it is pointed out that the controller need not actually perform measurements on the state, but locally preparing and randomly distributing four possible two-qubit states~\cite{tittel2001Experimentala,williams2019Quantum}.
	Note that the choice of measurement of Alice is public information, thus only one bit of information is kept secret for each run in the practical implementation.
	For the purpose of control, we here note that one fixed measurement setting is enough.
	
	In our scheme, we randomly (with equal probability) assign 0 and $\pi$ relative phase to the entangled qubits, which is equivalent to performing measurement $\sigma_{\textsf{X}}$ at the controller side, thus the source distributing a random mixture of states $\kt{\phi^{+}}$ and $\kt{\phi^{-}}$ to the end users.
	Note that the two states have an opposite correlation with respect to measurements $\sigma_{\textsf{X}}$, $\sigma_{\textsf{Y}}$, and the superposition $\sigma_{\textsf{X} \pm \textsf{Y}}$.
	The security analysis of EB QKD usually requires demonstrating violation of Bell-type inequalities (e.g. Clauser-Horne-Shimony-Holt (CHSH) inequality~\cite{clauserProposedExperimentTest1969}).
	Specifically, the CHSH parameter~\cite{clauserProposedExperimentTest1969} can be written as
	\begin{equation}
	\begin{aligned}
		S_{\text{CHSH}}&:=
		E(\sigma_{\textsf{X}},\sigma_{\textsf{X+Y}})
		+E(\sigma_{\textsf{Y}},\sigma_{\textsf{X+Y}})
		+E(\sigma_{\textsf{X}},\sigma_{\textsf{X}-\textsf{Y}})\\
		&-E(\sigma_{\textsf{Y}},\sigma_{\textsf{X}-\textsf{Y}}),
	\end{aligned}
	\end{equation}
	where
	\begin{equation}\label{eq:corr}
		E(\alpha,\beta):=p_{\alpha,\beta}^{++}+p_{\alpha,\beta}^{--}-p_{\alpha,\beta}^{+-}-p_{\alpha,\beta}^{-+}
	\end{equation}
	is the correlation function comprising the probabilities of four possible outcome combinations given the measurements $\alpha$ and $\beta$.
	The correlation function $E$ quantifies the correlation of the photon pairs, assuming that the coincidence from output "${++}$" and "${--}$" signify correlated; and "${+-}$" and "${-+}$" signify anticorrelated.
	$S_{\text{CHSH}}$ is upper-bounded at $2$ under the assumption of \emph{locality} in classical regime.
	However in quantum regime, entangled states are strongly correlated surpassing the locality restriction~\cite{hensenLoopholefreeBellInequality2015,*giustinaSignificantLoopholeFreeTestBell2015,*shalmStrongLoopholeFreeTest2015} with the upper-bound of $2\sqrt{2}$.
	Thus $S_{\text{CHSH}}>2$ ensures that the source is genuinely quantum (or more generally, no-signaling~\cite{barrettNoSignalingQuantum2005}) and not manipulated by attackers.
	
	For state $\kt{\phi^{+}}$ the expected value of $S_{\text{CHSH}}$ is $2\sqrt{2}$, while for $\kt{\phi^{-}}$ is $-2\sqrt{2}$.
	Thus the uniform mixture of $\kt{\phi^{+}}$ and $\kt{\phi^{-}}$ results in $S_{\text{CHSH}}=0$, if the two states are not properly identified.
	In other words, how well the two users know about the exact correlation of each state determines whether the violation can be certified.
	We leave the quantification of the effect of phase modulation to the subsequent section.
	
	The above scheme has one main technical problem: how to efficiently deliver the modulation information to the legitimate users. 
	In a faithful execution of the QSS protocol, the raw key held by the dealer corresponds to the modulation information, which has an equal size to the raw key held by each user.
	This forces the controller to actively participate in the implementation and record a phase bit for each coincidence event the users detect, thereby adding to implementation complexity.
	More preferably, the controller can preshare a string of modulation information with legitimate users and they select the bit which coincides with a photon detection.
	This however requires the string to be much longer than the final raw key to offset the significant loss \textit{en route}. 
	One may expect that repeating a fixed-length random binary string can work as the preshared modulation string.
	However, as we explained in Appendix~\ref{App1}, the unauthorized users can always collaborate and reveal the pattern given any finite-size preshared modulation string. 	
	As the source needs to generate random bits throughout the session, a practical solution is to use a block cipher algorithm [e.g. Advanced Encryption Standard (AES)~\cite{Daemen:2002:DRA,*AES-FIPS}] working in stream encryption mode, which is commonly used for bit expansion in prepare-and-measure QKD implementations~\cite{Walenta2014}.
	In this way only a small size of preshared randomness is required as the seed. 

\section{Genuine phase randomization and defense against memory attack}\label{Sec:mod2}
	The above scheme though prevents unwanted usage, cannot directly apply to the defense against memory attack~\cite{barrettMemoryAttacksDeviceIndependent2013}, in which the attacker is able to learn fully about the measurement choices and outcomes of users.
	If the attackers combine the data of key generation that was recorded in the previous session, as we discuss in Appendix~\ref{App1}, due to the correspondence of measurement results and modulation bits, they know immediately the modulation string according to the raw key.
	The question is whether it is possible to protect the modulation information from revealing even when the raw key is fully exposed.
	We show in the following that this is indeed achievable.
	
	The basic idea is to hide the detectable binary pseudo randomness in genuine randomness. 
	Operationally, we choose to apply a random phase with a probability of $p$
	to the maximally entangled states so as to destroy the phase correlation.
	This is practically equivalent to distributing a mixed state to the users, which can be written as 
	\begin{equation}
	\hat R=(\kt{0}\ba{0}\otimes \kt{0}\ba{0}+\kt{1}\ba{1}\otimes \kt{1}\ba{1})/2.
	\end{equation}
	Thus, the source now randomly distributes three possible states to the users: $\kt{\phi^{+}}$, $\kt{\phi^{-}}$ and $\hat R$ with respective probabilities of $(1-p)/2$, $(1-p)/2$ and $p$.
	With the local states at each user unchanged ($\mathbbm{1}/2$), the malicious devices cannot reveal the control string from the measurement outcomes because it is impossible to retrieve a specific form of decomposition out of a mixed state without sufficient ancillary information.
	Even if the devices recorded the raw key and managed to send it out, the eavesdropper still cannot obtain the secure key in previous sessions because the key generation process is encrypted.
	More precisely, the effective raw key is obtained by sifting out the genuinely random bits from the measurement results.
	We note that the measurement results should be transferred to and postprocessed on isolated and trusted computers which hold the modulation information.

\section{Controlled entanglement source protocol}
	We now give a detailed description of how our protocol proceeds.
	
	(1) Similar to traditional EB QKD protocols, the distributor and users reconcile the configuration of their respective setups including the synchronization of the clocks~\cite{fitzkeScalableNetworkSimultaneous2022}, but with an extra parameter, the token of the preshared seed.
	
	(2) The selected seed is locally extended to a continuous random string via the same block cipher algorithm working in stream encryption mode.
	Under a certain rule the (most likely binary) random string is transformed into ternary, in which the proportion of trit "2" is properly set.
	The rule can be quite simple.
	For instance, to check every two consecutive bits of the random string, one assigns "2" to the resulting ternary random string if the two consecutive bits are "11," otherwise assigns "00," "01," or "10," if "00," "01," or "10," respectively.
	According to the ternary random string, the controller constantly distributes one of the three element state $\kt{\phi^{+}}$ (trit "0"), $\kt{\phi^{-}}$ (trit "1"), and $\hat R$ (trit "2") to the users.
	
	(3) The users randomly pick measurement bases to their devices, and obtain a binary outcome ("+" or "$-$") upon a photon detection. Note that the measurement choice and the corresponding outcomes could be recorded and leaked if the memory attack is applied.
	
	(4) Given sufficiently many rounds, the controller stops distribution and does not take part in the subsequent procedures.
	Meanwhile the users publicly announce their measurement bases and sift out the unwanted combination of bases for the protocol.
	
	(5) Among the remaining results, the users choose to publicly announce part of the outcomes of the measurements that are related to security analysis, for instance, measurement $\sigma_{\textsf{X}}$, $\sigma_{\textsf{Y}}$, and $\sigma_{\textsf{X} \pm \textsf{Y}}$.
	The outcomes of each measurement combination form a classical bit-string $\bm{T}^{1(2)}_\text{O}$ at each side, where the superscripts denote the users.
	As the clocks are synchronized, a decoding string $\bm{T}_\text{D}$ can be deduced out of the local random string prepared in stage (1), according to the time of the photon detection---if the user is legitimate.
	By further omitting trit "2" in $\bm{T}_\text{D}$, and the corresponding bit in $\bm{T}^{1(2)}_\text{O}$, the security can be certified with either user computing the CHSH function from the statistics of $\bm{T}^{1}_\text{O} \oplus \bm{T}_\text{D}$ and $\bm{T}^{2}_\text{O}$, or $\bm{T}^{1}_\text{O}$ and $\bm{T}^{2}_\text{O} \oplus \bm{T}_\text{D}$.
	During this stage, unauthorized users will fail the test, as they will not hold a valid $\bm{T}_\text{D}$.
	
	To have this more concrete,	it is convenient to assume that the user actually holds a string $\bm{T}_\text{U}$, and we define the effectiveness of control using the correlation between $\bm{T}_\text{U}$ and $\bm{T}_\text{D}$:
	\begin{equation}
		U:=\sum_{j\in \bm{T}}^{N}\frac{(-1)^j}{N},
	\end{equation}
	where $\bm{T}=\bm{T}_{\text{U}} \oplus \bm{T}_{\text{D}}$ with trits "2" omitted, and $N$ is the length of $\bm{T}$.
	Note that the correlation of mixed state $\hat R$ is always zero, and the probability of picking entangled elements is $(1-p)$.
	Thus with modulation the correlation function can be written as
	\begin{equation}\label{eq:Emod}
		E_{\text{mod}}=U(1-p)E,
	\end{equation}
	where $E$ is defined in Eq.~\eqref{eq:corr}.
	For example if $T_{\text{D}}=``01010101,"$ and $T_{\text{U}}=``00000000,"$ then $U=0$.
	In such a case, the users cannot distinguish completely the correct correlation that should be adopted, which will result in $E_{\text{mod}}=0$, therefore the obtained $S_{\text{CHSH}}=0$ (after modulation), and they will have to abort to avoid security risks.
	For a general $T_{\text{D}}$, in Appendix~\ref{App2} we derive the optimal $U$ for unauthorized users, which is far below the requirement for demonstrating inequality violation, given a reasonably long modulation string.
	
	(6) Once the security is certified, the users can implement the conventional error correction and privacy amplification method to obtain the final secure key, based on the outcomes that are acquired in the same measurement bases. 
	This however means that if the outcomes are exposed, the attacker can always generate the identical key as the method of error correction and privacy amplification are tacitly open, which is exactly how memory attack breaches.
	By introducing the mixed state, a certain proportion of the raw outcomes are noncorrelated and are omitted at both sides in the postprocessing stage, which is unknowable to the attackers.
	Thus the secure key deduced from the raw outcomes at the attacker's side is noneffective, despite the same correction and amplification method.
	The proportion of the mixed state, however, comes with a trade-off, since the mixed state carries no information but is to interfere the attacks.
	We note that the setting of $p$ varies depending on the practical situation, such as the channel loss.
		
	We acknowledge that the security against memory attack ultimately derives from the classical secret and the QKD protocol produces no extra permanently secure data.
	Thus we emphasize that to ensure the long-term reliability of the protocol, the controller needs to communicate regularly (say daily with a small fraction of time) with other users for refreshing the seed rather than using a fixed preshared seed.
	It is worth mentioning that the controller though controls the QKD implementation, does not gain any information about users' secure key.
	
\begin{figure}[ht]
		\includegraphics[width=\columnwidth]{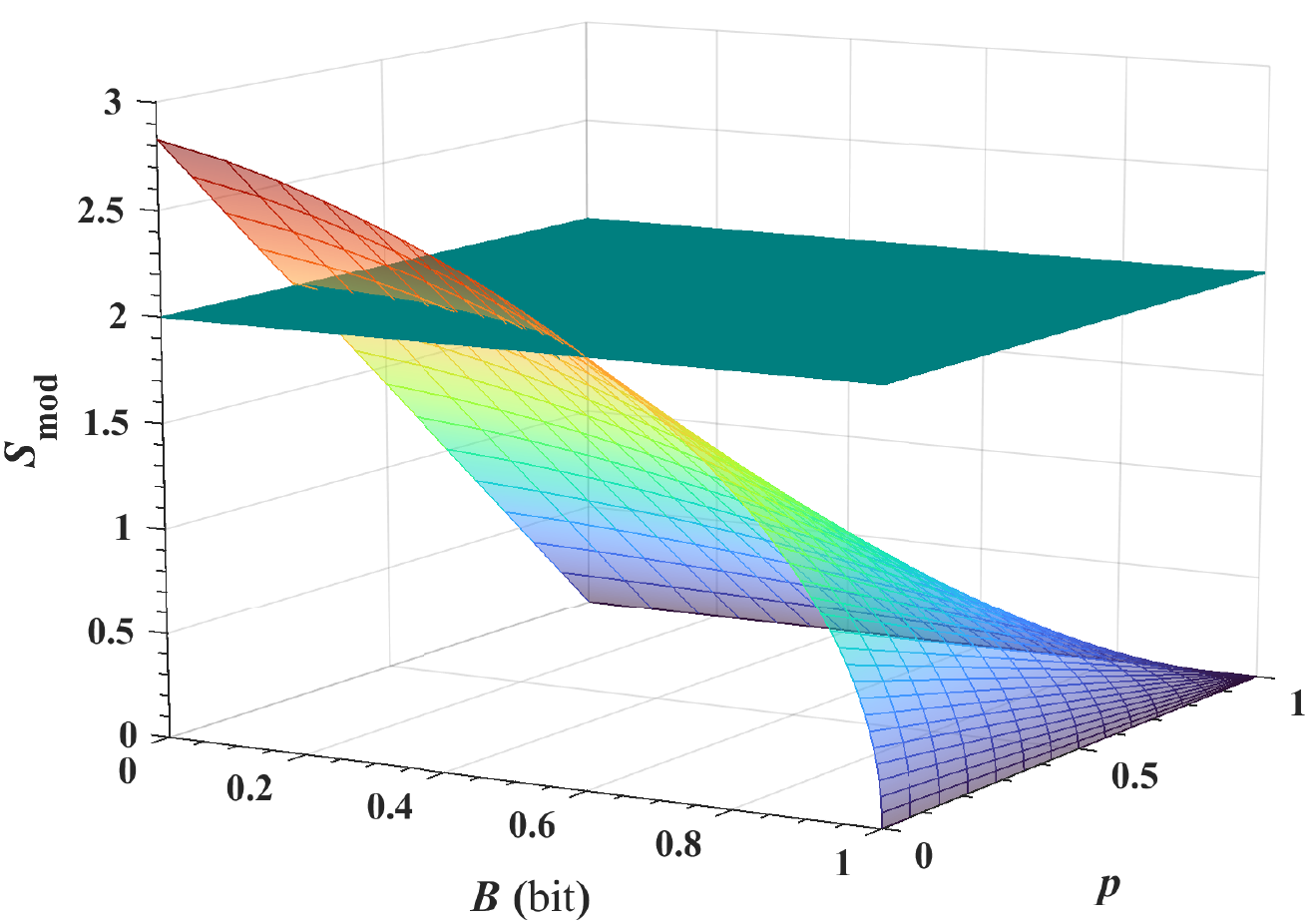}
		\caption{\label{F2}
			Diagram of $S_{\text{CHSH}}$ being a function with respect to $B$ the binary entropy and $p$ the proportion of the mixed state.}
	\end{figure}

	We name the above scheme \emph{controlled entanglement source (CES)} protocol, under which the estimated relation of $S_{\text{mod}}$ versus $U$ and $p$ is
	\begin{equation}\label{eq:Smod}
		S_{\text{mod}}=2\sqrt{2}(1-p)U,
	\end{equation}
	which is a direct deduction from Eq.~\eqref{eq:Emod}.
	As in the practical implementation, one cares only about the absolute value of $S_{\text{CHSH}}$, we introduce $B=h((1+U)/2)$ to reduce $S_{\text{mod}}$ to positive axis for simplicity, where
	$h(x)=-x\log_2(x)-(1-x)\log_2(1-x)$ is the binary entropy function.
	Note that $B$ \textit{per se} has a clear physical meaning, that is, the amount of uncertainty that is to be eliminated for the user to retrieve the correlation.
	The expression of $S_{\text{mod}}$ is visualized in Fig.~\ref{F2} forming a curved surface.
	The green plane denotes the threshold of CHSH violation, and the intersection indicates how much information (or how well the mixed state can be distinguished) is needed for the users to demonstrate the violation.
	For instance, the violation is possible should the entropy $B$ is less than 0.6 bit, according to the intersection point in the ($x,z$) plane (corresponding to no mixed state is involved), which is equivalent to that over 85.3\% of users' decoding string is identical to the modulation string (or its complement).
			
	\begin{figure}[ht]
		\includegraphics[width=1\columnwidth]{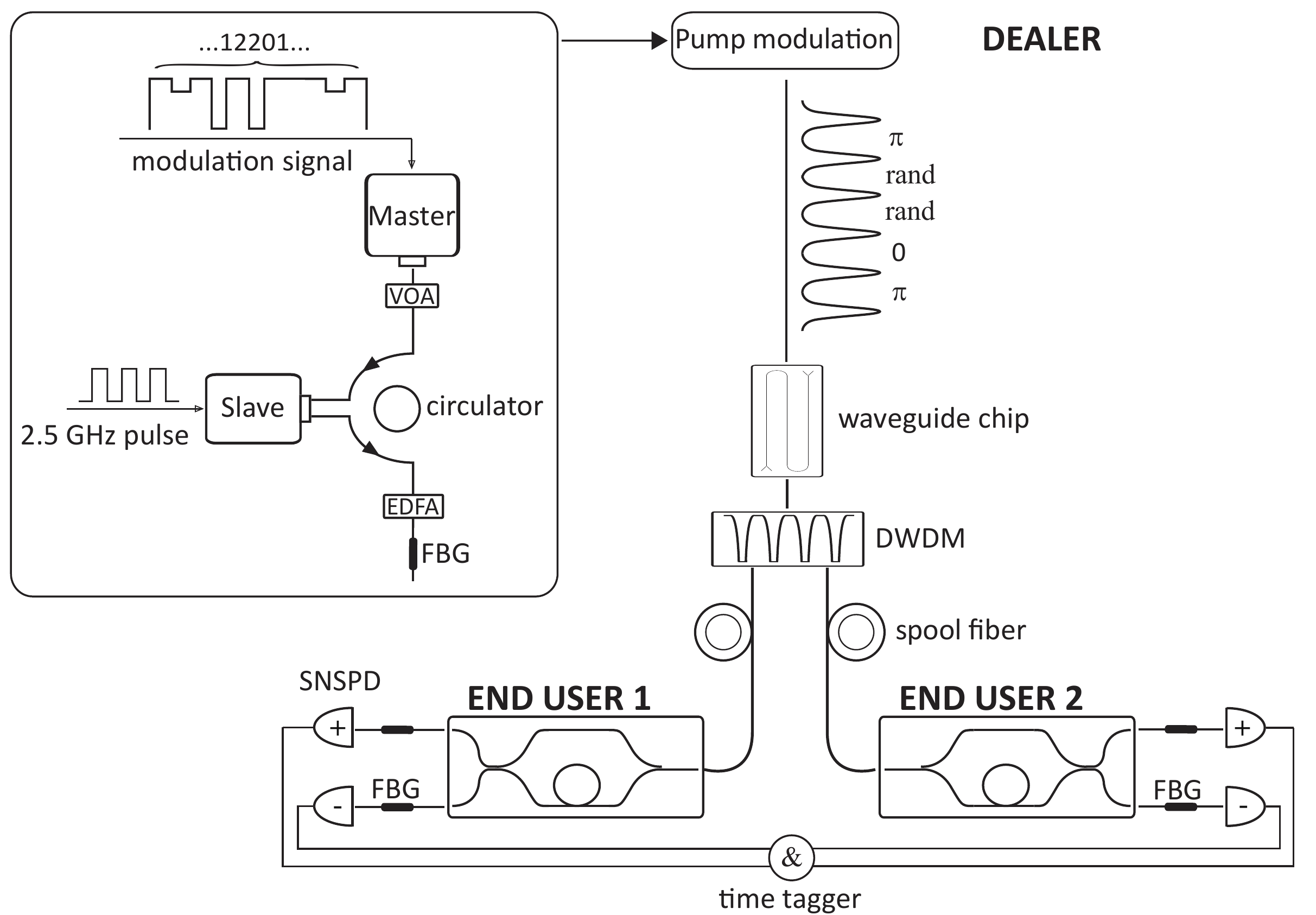}
		\caption{\label{F3}
			Experimental setup of controlled entanglement source.
			Note that digit "2" in the modulation string refers to genuine randomized phase.
			VOA, variable optical attenuator;
			EDFA, erbium doped fiber amplifier; 
			FBG, fiber Bragg Grating; 
			SNSPD, superconducting nanowire single-photon detector.
		}
	\end{figure}	
	
\section{Experimental setup and results}\label{Sec:exp}
	To demonstrate the feasibility of our protocol, we implement an experimental setup using off-the-shelf components as schematically shown in Fig.~\ref{F3}.
	At the dealer's side, a train of phase-modulated~\cite{yuan2016Directly} pump pulses is sent to a silicon waveguide chip~\cite{llewellynChiptochipQuantumTeleportation2020,chenQuantumEntanglementPhotonic2021}, in which the time-bin entanglement states are realized through spontaneous four-wave mixing (SFWM) effect.
	Note that in our setup, the asymmetric interferometer which is traditionally deployed at the pump side to induce time-bin superposition state~\cite{tittel2001Experimentala} is reduced, which implies that the delay of time bins of our source is reconfigurable.
	The resulting broadband signal-idler spectrum is then directed to a dense wavelength division multiplexer (DWDM) module thus being divided into multiple channels.
	Here channels C30 and C37 of the International Telecommunication Union (ITU) grid (full width at half maximum: 25~GHz) are selected for demonstration, and we note that more channels are directly supported in our system.
	The fiber spool at each quantum link is simulated with a 5-dB fixed attenuator in this proof-of-principle demonstration.
	Finally the end users receive the controlled entangled states to implement quantum cryptography.
	A detailed description of the generation of modulation signal and correlation test is provided in Appendix~\ref{App:exp}.

\begin{figure}[ht]
		\centering
		\subfloat{
			\includegraphics[width=\linewidth]{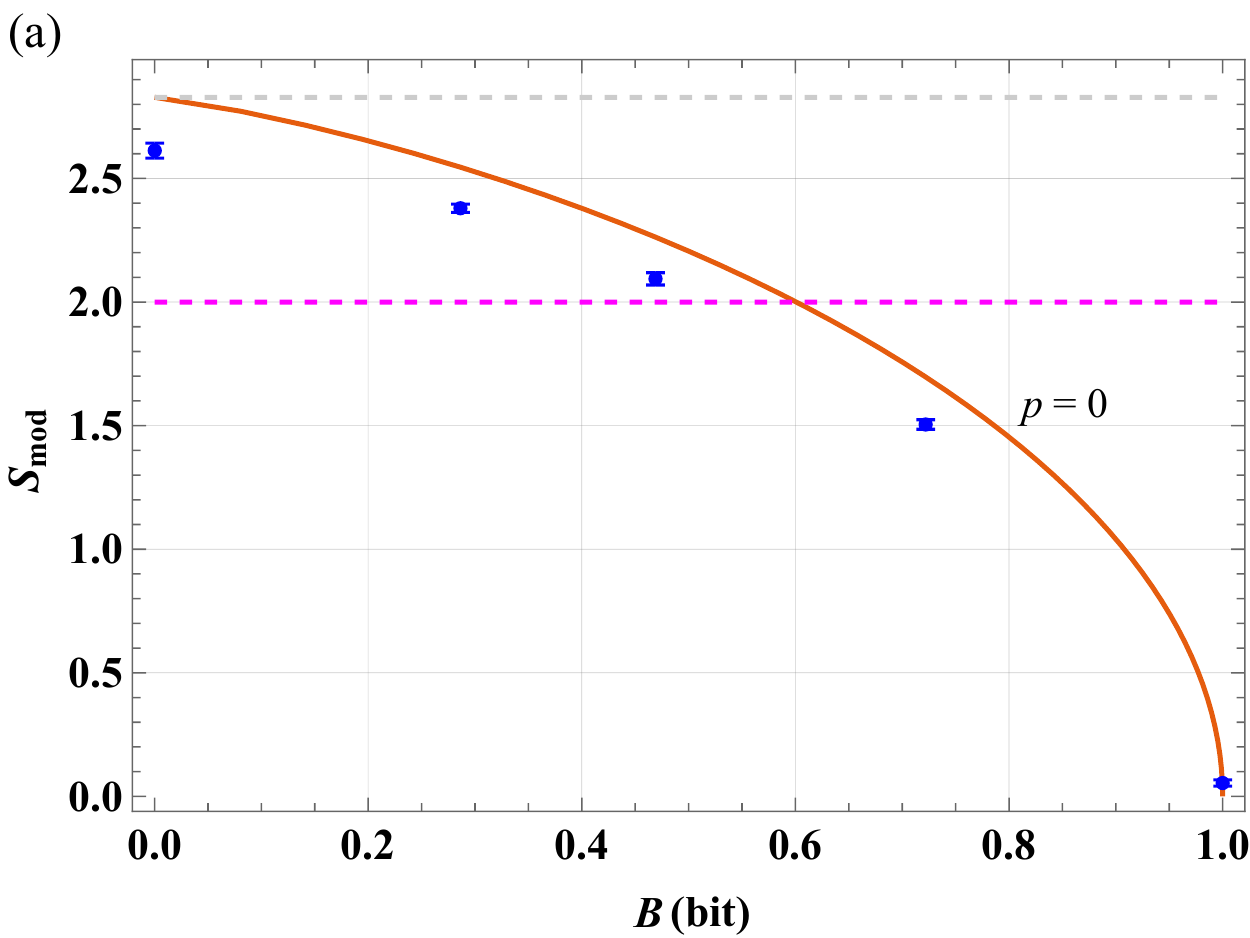}\label{F4a}
		}
		\newline
		\subfloat{
			\includegraphics[width=\linewidth]{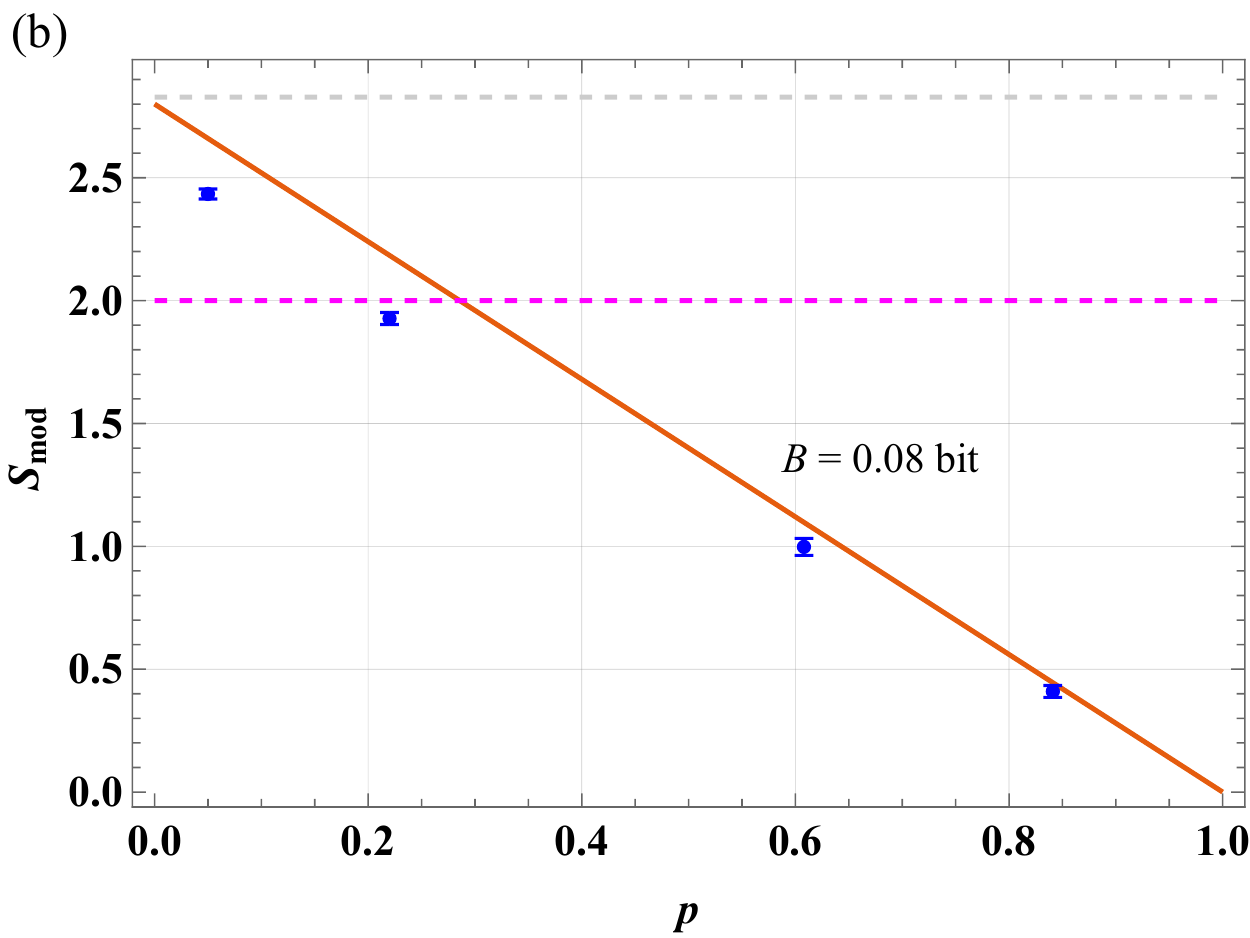}\label{F4b}
		}
		\caption{Experimental results of CHSH parameter versus different controlling configurations. 
			Red solid lines denote the theoretical prediction, and blue dots denote the experimental results.
			Magenta dashed lines denote the threshold for inequality violation.
		}
	\end{figure}

	The experimental results are as follows: the overall heralding efficiency of the experimental setup is 1.6\%, including losses from the photonic waveguide chip ($-6.5$ dB), transmission and coupling ($-7$ dB), spectral filtering ($-4$ dB), and photon detection ($-0.5$ dB).
	The observed maximal coincidence rate exceeds 2 kHz, including the ones detected in time basis (the side peaks), with a coincidence to accidental ratio (CAR) of 50.
	In Fig.~\ref{F4a} and~\ref{F4b} we present the modulated CHSH parameter results, which are in good agreement with the theoretical model.
	The gap between data and theory arises mainly from the imperfect interference, and the observed entanglement visibility is 96.1(5)\% (see Appendix~\ref{App:interference} for more results), which \textit{per~se} well suffices the requirement of defending against collective attack in device-independent scenario~\cite{pironioDeviceindependentQuantumKey2009}.
	Note that collective attack is independent of other attacks.
	
\section{Conclusion}\label{Sec:Con}
	Control of access to entanglement resource is shown to be critical in preventing quantum cryptography from unwanted usage and malicious attacks.
	Inspired by the quantum secret sharing protocol, we propose a protocol to endow conventional entanglement sources with access control, by introducing a twofold phase randomization.
	We have shown that by employing our protocol under onefold phase randomization, the entanglement distributor can effectively control users.
	Further, memory attack on measurement apparatuses can be defended against with the protocol under twofold phase randomization, and the cost is negligible as the measurement devices can be safely reused in our protocol.
	
	We experimentally verify the feasibility of our protocol in a time-bin encoded system using off-the-shelf devices, in which the control to entanglement source is precisely demonstrated.
	In our setup, the interferometer at the pump side is reduced, which significantly improves the robustness and grants reconfigurability of the delay of time bins to the source.
	The observed entanglement visibility of our setup is over 96\%, which indicates that the traditional collective attack is individually addressed.
	We believe an upgrade of current entanglement source bases on our protocol can significantly improve its performance and flexibility in quantum cryptography applications.
	
	\begin{acknowledgments}
	The authors thank Professor Wei Zhang for providing Si waveguide chips in the early stage of the investigation.
	This work is supported by National Natural Science Foundation of China under Grants 62105034 (L. Z.), 12105010 (Q. Z.), and 62250710162 (Z. Y.). 
	\end{acknowledgments}
	 
	\appendix
	
\section{Cheating scheme for unauthorized users} \label{App1}
	We first note that in the original three-party QSS protocol, only one of the participants is assumed to be dishonest and the dealer will share its information with the honest one.
	Notably, there is no point to QSS if both participants are dishonest.
	In a controlling scenario, the two unauthorized users both gain no information from the source, but it is possible for them to infer which state they are indeed sharing if they collaborate (as they definitely will), which is distinct from QSS but rather fits the third-man quantum cryptography scenario.
	Trivially, for state $\kt{\phi^{+}}$, they will with almost certainty (depending on the entanglement visibility) observe coincidence event at the port "++" or "$--$" if they both measure $\sigma_{\textsf{X}}$.
	With the same measurement setting, if they are alternatively sharing state $\kt{\phi^{-}}$, they will very much likely observe a coincidence event at the output "$+-$" or "$-+$".
	Thus given sufficiently many events, they can finally find the pattern of the modulation string no matter how long (finite) it is.
	
	The only limitation is that on revealing modulation string, they cannot simultaneously do the security certification, which requires switching nonorthogonal measurement bases; nor generate secret keys, where communicating measurement outcomes is prohibited.
	Hence, the source needs to distribute random bits throughout the session rather than repeating a fixed-length random string, to make sure the unauthorized users do not exploit a pattern to implement QKD protocol. 
				
	\begin{figure}[ht]
		\includegraphics[width=\columnwidth]{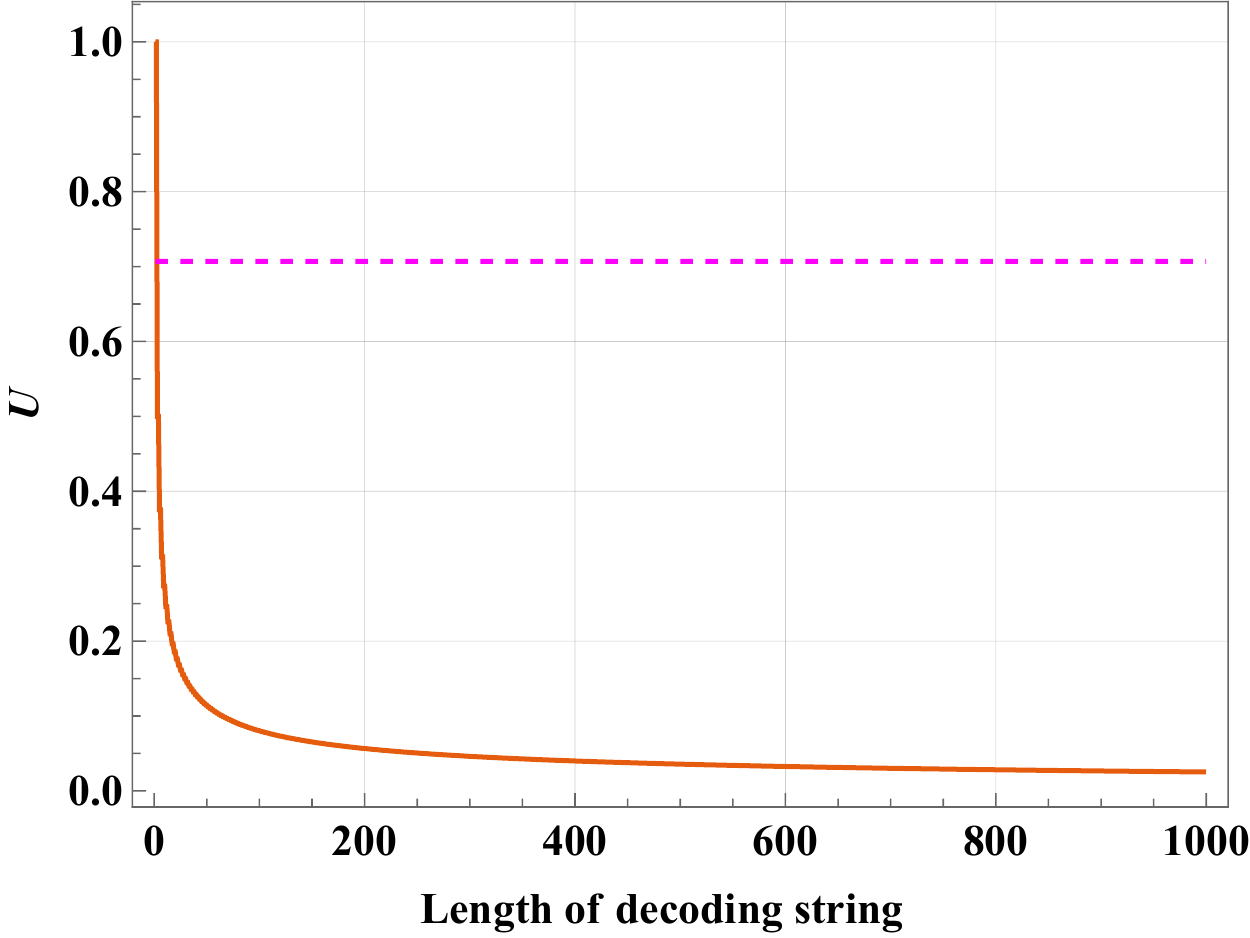}
		\caption{\label{AF1}  
			The expected value of $U$ for unauthorized users as a function of $N$.
			Magenta dashed line denotes the threshold for inequality violation.
		}
	\end{figure}

\section{Optimal $U$ for unauthorized users} \label{App2}
	Without ancillary information, the unauthorized users can only guess a $\bm{T}_{\text{D}}$.
	For simplicity, we assume that $\bm{T}_{\text{D}}$ comprises fully "0".
	If for example, $\bm{T}_{\text{D}}$ and $\bm{T}_{\text{U}}$ have only one bit difference (specific location of the different bit does not matter), then $U=(N-2)/N$ as there are two opposite bits and would cancel each other. 
	Generally we derive that 
	\begin{equation}\label{Aeq1}
		U=\frac{1}{2^{N-1}N}\sum_{i=0}^{\lfloor N/2 \rfloor}\binom{i}{N}(N-2i),
	\end{equation}
	the expected value for unauthorized users.
	Note that the probability of "$\bm{T}_{\text{D}}$ and $\bm{T}_{\text{U}}$ have $n$ bit difference" is $\frac{1}{2^{N}}\binom{n}{N}$, and we only consider positive $U$ here.
	In Fig.~\ref{AF1} we present $U$ as a function of $N$, in which one finds that $U=1$ thus above the threshold $1/\sqrt{2}$ only for $N=0,1$.
	For a reasonably long modulation string, it is impossible for unauthorized users to reproduce $\bm{T}_{\text{D}}$ and demonstrate violation of Bell-type inequalities.

\section{Generation of modulation signal and the correlation test} \label{App:exp}
	As the CES protocol requires three element states, we adopt the direct phase modulation technique~\cite{yuan2016Directly} to accurately introduce the desired phase shift or phase-randomization to the time-bin entanglement states.
	A pair of gain-switched distributed feedback (DFB) lasers (Gooch \& Housego, typical linewidth, 1 MHz; pulse duration, 80 ps) are modulated by 2.5-GHz signals and employed in optical injection-locking configuration. 
	The remarkable advantage of this setup is that the phase randomization can be readily achieved by switching the master driving signal across the master laser's lasing threshold.
	Operationally, as illustrated in the inset of Fig.~\ref{F3} in the main text, state $\kt{\phi^{+}}$ corresponds to applying a regular driving voltage to the master laser (signal "0"); state $\kt{\phi^{-}}$ corresponds to applying a shallow driving voltage (signal "1"), while state $\hat R$ corresponds to a reversing voltage (signal "2").
	The master laser centered at 1550.52 nm has optical power approximately 80$~\mu \rm W$ varying depending on the modulation signal.
	A 2.5-GHz square signal is applied to the slave laser for pulse preparation.
	The output power of EDFA is $10~\rm mW$.
	
	For the detection part, two AMZIs are placed, respectively, at the two end users, which can control the phase thus setting desired measurement bases by varying the temperature.
	The AMZI is homemade, which comprises two polarization-maintained 50:50 fiber couplers and a set of temperature control modules, and is enclosed by a box to shield it from environmental temperature fluctuations.
	The time delay of the AMZI is 400 ps, which is in line with the pump-pulse repetition frequency.
	The phase of AMZI is actively stabilized based on temperature, and a cw reference light beam (not shown in the figure) centered at 1550.52 nm is used to provide the feedback control.
									
	\begin{figure}[t]
		\includegraphics[width=\columnwidth]{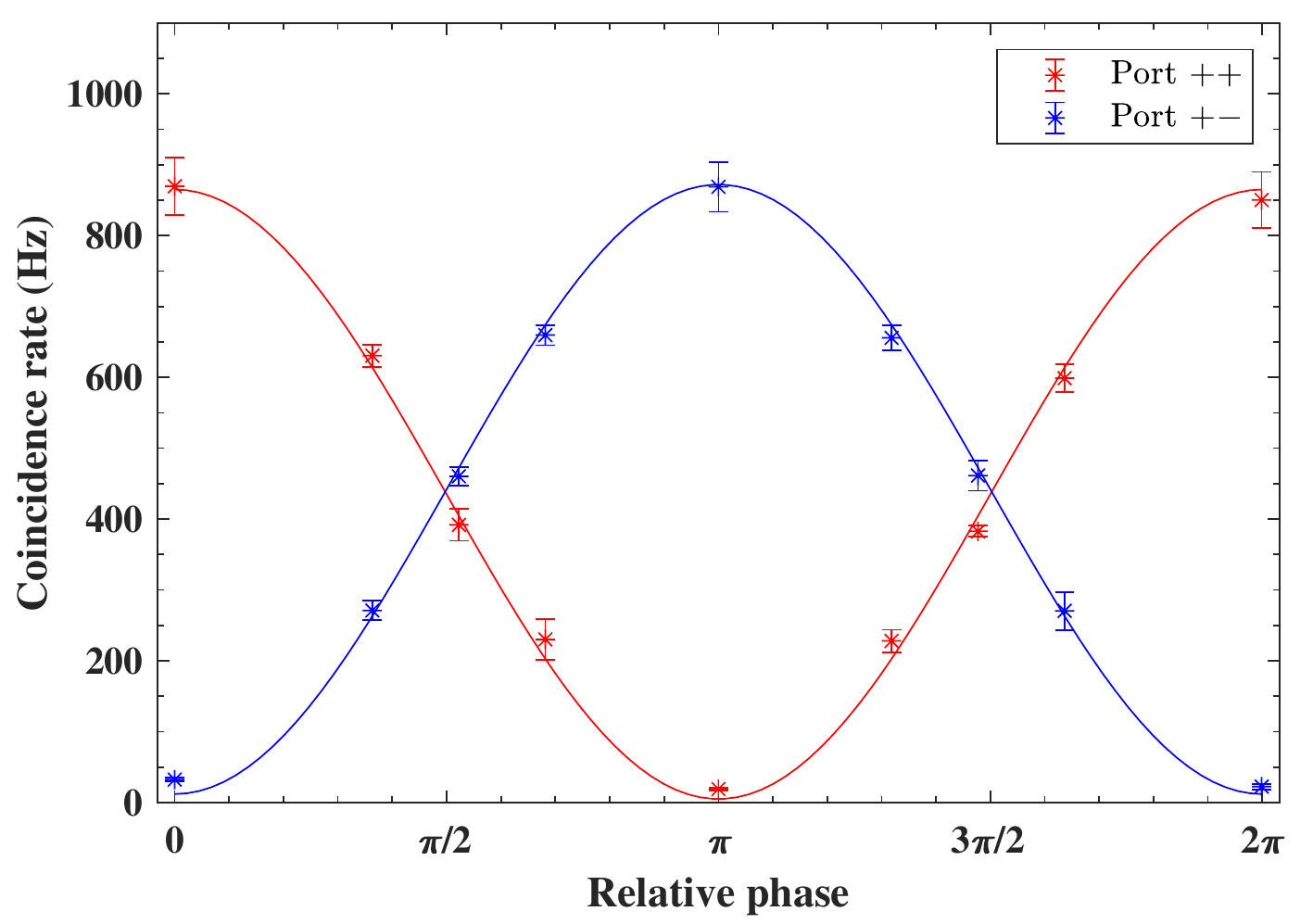}
		\caption{\label{AF2}
			Interference fringes of AMZI with respect to $\kt{\phi^{+}}$.}
	\end{figure}
	
	To simulate the cases where unauthorized users or malicious devices somehow gain partially the modulation information, we deliberately set several nontrivial modulation configurations, and then implement the CHSH test.
	That is to say, we prepare beforehand several nonuniformly distributed random strings such that there is a "bias" in the modulation which can be regarded as the consequence of breaching.
	In the postprocessing procedure, we do not implement any operation to the outcome string, this is equivalent to setting a plain $\bm{T}_\text{U}$ (with full "0" or full "1"), and thus $U$ is solely dependent on $\bm{T}_\text{D}$ in our experiment.
	Through the above settings, we essentially characterize how $S_\text{mod}$ varies given designated values of $B$ and $p$.
	Note that in the practical implementation, we first obtain the value of $U$, then we compute $B$.
	The generation of the modulation strings is accomplished by using random number generator on a desktop computer. 
	For this proof-of-principle demonstration, instead of a large continuous random string via using block cipher algorithm, we adopt a random string comprising 5000 random classical trits, and then load the sequences of string on an arbitrary wave generator (AWG, Tektronix, 70002B) and repeat it to produce the modulation signal.
	At the users' side, photon coincidence is registered using a time tagger (ID Quantique, ID900) for four pairs of output simultaneously with the time bin of width 100 ps.
	For each pair of outputs, there are three possible photon arrival times which results in three equidistant coincidence peaks in the arrival-time histogram.
	The coincident events in the central peak yield detections in the phase basis, according to which we compute the correlation $E$ and hence the modulated CHSH parameter $S_{\text{mod}}$ by varying the measurement basis.

\section{Characterization of interferometer} \label{App:interference}
	In Fig.~\ref{AF2} we present the experimental results of biphoton interference fringes of outputs "$++$" and "$+-$" of the one of the AMZIs, when the source is fixed with state $\kt{\phi^{+}}$.
	The red dots denote the rate of coincidence counts from output "++", and blue dots denote the rate of coincidence counts from output "$+-$".
	The solid curves are the fitted sinusoidal function of the photon coincident rates.


\providecommand{\noopsort}[1]{}\providecommand{\singleletter}[1]{#1}%

\end{document}